# On Optimizing Locality of Graph Transposition on Modern Architectures

Mohsen Koohi Esfahani

Hans Vandierendonck

*Abstract*—This paper investigates the shared-memory Graph Transposition (GT) problem, a fundamental graph algorithm that is widely used in graph analytics and scientific computing.

Previous GT algorithms have significant memory requirements that are proportional to the number of vertices and threads which obstructs their use on large graphs. Moreover, atomic memory operations have become comparably fast on recent CPU architectures, which creates new opportunities for improving the performance of concurrent atomic accesses in GT.

We design PoTra, a GT algorithm which leverages graph structure and processor and memory architecture to optimize locality and performance. PoTra limits the size of additional data structures close to CPU cache sizes and utilizes the skewed degree distribution of graph datasets to optimize locality and performance. We present the performance model of PoTra to explain the connection between cache and memory response times and graph locality.

Our evaluation of PoTra on three CPU architectures and 20 real-world and synthetic graph datasets with up to 128 billion edges demonstrates that PoTra achieves up to 8.7 times speedup compared to previous works and if there is a performance loss it remains limited to 15.7%, on average.

## I. INTRODUCTION

Graph Transposition (GT) is a fundamental graph algorithm that plays a crucial role in various graph analytic algorithms and is also utilized for transposing matrices in scientific computing. GT applications include analyzing traffic networks, gene expression, ranking proteins in biological networks, simulation of navigation systems, and studying social networks.

The long execution time of graph transposition and emergence of trillion-scale real-world public graphs [1]–[5] have made efficient GT a challenge. The previous attempts to optimize GT have been mainly based on the assumption that *concurrent atomic memory accesses cannot be performed efficiently*. Based on this assumption, the state-of-the-art GT algorithms [6] try to avoid atomic memory accesses. However, these algorithms require **per-thread data structures that grow linearly with the number of vertices**. This limitation restricts the size and order of the input graph, negatively impacting three key performance metrics: parallelism, cache locality, and load balance.

Moreover, with the advent of new CPU architectures, the assumption that atomic accesses are inefficient no longer holds true. We show that, in contrast to a few processor generations ago, **current CPUs are capable of performing atomic memory accesses at comparably close rates to non-atomic accesses**. This presents an opportunity to design algorithms that leverage the good performance of atomic primitives while avoiding excessive memory demands.

Our evaluation reveals that the primary cause of prolonged execution time in GT is the use of shared arrays between threads, which leads to cache contention. Furthermore, the cache is unable to accommodate these shared arrays due to its limited capacity, resulting in a high cache miss rate. This shows that the **GT problem needs new solutions to address the locality of memory accesses without imposing restrictive memory space requirements**.

In this paper, we consider the GT challenge from a novel perspective: graph structure. Large-scale real-world graphs such as those derived from social networks, internet web pages, and bioinformatics, exhibit a **skewed degree distribution**. We leverage this characteristic to categorize vertices and optimize performance. The skewed distribution of degrees means that Low-Degree Vertices (**LDV**) form a large portion of vertices with each having only a few connections. On the other hand, High-Degree Vertices (**HDV**) are rare and fewer in number, but each HDV has a large number of edges.

We optimize the locality of memory accesses by considering **the implications of the graph structure on GT execution**. We distinguish the way LDV and HDV edges are processed. Since LDV are frequent but rarely referenced, it is inefficient to allocate per-thread memory for them or keep them in the CPU cache. In contrast, HDV are few in number but frequently accessed by their neighbors, making them suitable candidates for hosting in the CPU cache.

Based on this idea, we design the **PoTra** algorithm as a structure-aware GT algorithm that optimizes locality and performance without requiring large-size auxiliary data structures. PoTra employs a hash table to compress vertex IDs of HDV into consecutive indices of per-thread arrays, whose total size is less than the size of the CPU caches. Furthermore, to leverage the capabilities of recent CPU architectures, PoTra increases atomic memory accesses for graphs with a good locality level.

To gain a deeper understanding, we present a performance model of PoTra that links its effectiveness to the response time of different memory access types and graph locality.

We evaluate PoTra in comparison to state-of-the-art GT on three CPU models (AMD Zen2 and Zen3 and Intel Sapphire Rapids) and using 20 real-world and synthetic graph datasets with up to 128 billion edges. The evaluation shows that PoTra achieves up to 8.7× speedup compared.

The contributions of this paper are:

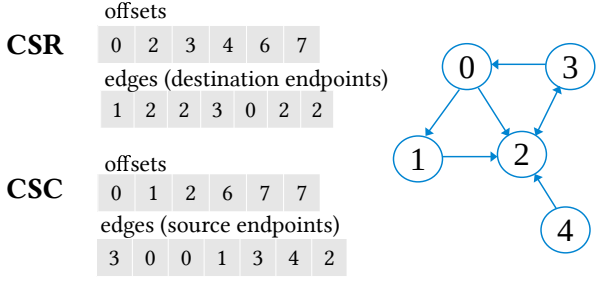

Fig. 1: CSR and CSC presentations of a sample graph

- An analysis of the performance of atomic and non-atomic memory access types on CPU architectures and its implications on GT,
- A novel method for designing architecture-aware algorithms that optimally match the dimensions and latency characteristics of architecture components,
- The design of PoTra, a novel structure-aware GT algorithm that optimizes locality and performance,
- Modeling performance of different GT algorithms, and
- Evaluation of PoTra in comparison to previous works.

This paper is structured as follows. Section II overviews the terminology and baseline GT algorithms. Section III motivates the discussion. Section IV presents the design of the PoTra algorithm and its performance model. Section V evaluates PoTra. Related works are reviewed on Section VI and Section VII concludes the discussion.

## II. BACKGROUND

In this section, we review the terminology and state-of-the-art GT algorithms.

### A. Terminology

Graph G = (V, E) consists of a set of vertices V, and a set of directed edges E with E ⊆ V × V. G is undirected if for each edge (u, v) ∈ E; (v, u) ∈ E. Otherwise, the graph is directed.

The Compressed Sparse Row (CSR) or Column (CSC) [7], [8] format is a compact representation for sparse matrices and graphs. It represents graphs using two arrays:

- the **offsets** array containing |V| + 1 elements, and
- the **edges** array of |E| elements.

The offsets array is indexed by a vertex ID and specifies the index of the first edge of that vertex in the edges array. The edges array specifies the ID of the *source* endpoints of the edges in CSC and *destination* endpoints in CSR.

This way, the CSR format lists the *outgoing* edges of each vertex and the CSC format lists the *incoming* edges of each vertex. In undirected graphs, each edge represents a bi-directional connection between two endpoints, making the CSR and CSC presentations similar.

Figure 1 shows a sample graph and its CSR and CSC presentations. In the CSR format, the offsets array starts with values 0 and 2, indicating that the outgoing edges of vertex 0 begin at index 0 and continue up to index 2 in the edges array. This means vertex 0 has outgoing edges to 1 and 2. Since vertex 0 has one incoming edge from vertex 3, in the CSC format, the offsets array starts with values 0 and 1, and the edges array begins with value 3.

*Graph transposition* transforms the CSC presentation of the graph to CSR or the CSR to CSC. The transposed of a transposed format is equal to the original format. In *edge-weighted graphs*, each edge in the CSR format is an ordered pair (ID, weight), where ID is the ID of the endpoint vertex and weight is the weight/label of the edge. Since the absence or presence of edge weights does not affect the GT algorithms, without loss of generality and for the sake of simplicity, we limit our discussion to unweighted graphs.

### B. Atomic Transposition

Algorithm 1 presents the pseudo code of atomic transposition, which consists of three steps:

**Step 1: Calculating Degrees**. The first pass over edges is performed to identify the degree of each vertex in the transposed graph. fetch_and_add() atomically increments the relevant indices. Note that each edge in G.edges stores only one endpoint, i.e., a vertex ID (Section II-A).

**Step 2: Calculating Offsets**. By the end of Step 1, the degree of all vertices has been computed. The t_offsets array (i.e., the offsets array of the transposed graph) is computed in parallel using the prefix_sum() function. The t_offsets array is used as the **Insertion Points (IP)** array in the next step[1].

**Step 3: Writing Edges**. The second pass over the edges is performed by processing all neighbors of each vertex. To determine the insertion point of neighbor u of vertex v, fetch_and_add() atomically increments IP[u] and returns its previous value, which is used as index to the

---

**Algorithm 1:** Atomic GT

**Input:** G(offsets, edges)

/* **Step 1: Calculating degrees** */
1 counters[|G.V|] = {0};
2 **par_for** e ∈ G.edges
3     fetch_and_add(counters[e], 1);
/* **Step 2: Calculating offsets and IP** */
4 t_offsets = prefix_sum(counters);
5 IP = t_offsets.copy();
/* **Step 3: Writing edges** */
6 t_edges[|G.E|];
7 **par_for** v ∈ G.V
8     **for** u ∈ v.neighbors **do**
9         index = fetch_and_add(IP[u], 1);
10         t_edges[index] = v;
11 **return** (t_offsets, t_edges);

---

[1]Actually, it is not required to have a copy of t_offsets as IP because, by completing Step 3, IP[i] becomes equal to t_offset[i+1]. Therefore, a single shift left for consecutive indices is sufficient to set IP to its initial value. The shift operation is performed in parallel across partition of vertices.

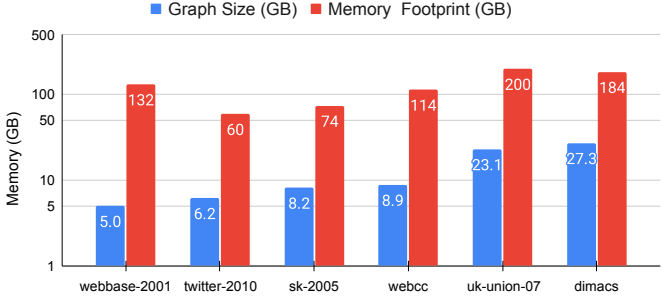

Fig. 2: [Zen2, CSR] Memory footprint of ScanTrans

t_edges array (the edges array of the transposed graph) to write v. By incrementing IP[u], the algorithm ensures that the next edge of u (if any) is written in the next index of the t_edges array.

An additional sorting step may be applied to sort the neighbor lists of each vertex in the transposed graph.

**Memory Access Types.** In Step 1 of Atomic GT, there are two types of memory accesses:

- Memory accesses for reading edges, which are performed **sequentially with no repeated access** to a memory address and benefit from hardware prefetchers, and
- Memory accesses for incrementing indices of the counters array, which are **random and are repeated** for all neighbors of a vertex.

The random accesses are the performance bottleneck of Step 1 due to their inefficient use of the memory system. Similarly in Step 3, random accesses to IP and t_edges arrays are the bottleneck.

### C. ScanTrans & MergeTrans

Atomic memory accesses may incur high performance overheads in Steps 1 and 3 of Atomic GT. To mitigate this, the ScanTrans algorithm [6] employs per-thread counters and IP arrays. In Step 1, each thread increments its private counters array which are then aggregated in Step 2 to create private IPs used in Step 3. It is crucial that all edges are processed by the same thread in Steps 1 and 3 to avoid mismatched IP.

MergeTrans [6] takes a different approach by dividing the graph into smaller subgraphs that fit within the cache during processing. Each subgraph is transposed in serial by a thread, and then pairs of transposed subgraphs are repeatedly merged.

Notably, both MergeTrans and ScanTrans do not require an additional sorting step. ScanTrans achieves sorted neighbor lists by assigning a single set of consecutive vertices to each thread. MergeTrans, on the other hand, sequentially reads and merges the edges arrays of two subgraphs during the merge phase, resulting in sorted neighbor lists.

## III. MOTIVATION

### A. Excessive Memory Demand

The MergeTrans and ScanTrans algorithms are designed under the assumption that atomic memory accesses are relatively slow. To address this, they replace atomic memory accesses with non-atomic accesses to thread-local copies of the counters array. While these algorithms are efficient for small graphs, their memory requirement limit their scalability to larger graphs (Section V-B). This is because **MergeTrans and ScanTrans rely on per-thread arrays resulting in a total memory space requirement of** $O(\#threads \cdot |V|)$.

Figure 2 illustrates the memory footprint of ScanTrans in comparison to the size of the graph (See Section V-A for the details of the graphs and the environment setup). The figure shows that memory footprint is 6–26 times of the graph size. This shows that **performance optimizations can become scalability bottleneck**, in this case, due to excessive memory requirements, which are dependent on both the input size and the number of threads.

### B. Atomics Are Not That Bad Anymore

The conventional wisdom that atomic accesses are expensive and should be minimized has been challenged by the evolution of processor architectures. Figure 3 presents a comparison of the execution rates of random "Write" and "Atomic Write" accesses under two conditions: (i) when the access is a hit in the L3 cache, and (ii) when access is a miss for the L3 cache. Each rate is normalized to random "Read" access rate when accesses hit in the L3 cache. The evaluation was performed on four CPU models: (1) SkyLakeX 6130, (2) CascadeLake W2295, (3) Sapphire Rapids 6438Y+, (4) Zen2 7702, and (5) Zen3 7773.

The evaluation involved measuring accesses of parallel threads (1 thread per core) to indices of an array. To measure the L3 hit rate, the total array size was set to the total size of the L3 caches in the machine. For L3 miss accesses, the array length was set to $1000\times$ the total L3 size. It should be noted that **accesses were random and not sequential**, aiming to have similar conditions in GT atomic access (Section II-B, Steps 1 and 3). The "xoshiro" pseudo random number generator[2] [9] was used to create random numbers. The source code of this benchmark is publicly available[3].

Figure 3 reveals that the performance degradation from "Write" to "Atomic Write" is significant on older architectures, 2.7 times degradation in SkyLakeX and 2 times in Cascade Lake. However, on newer architectures such as Sapphire Rapids, Zen2, and Zen3 atomic memory accesses are less than 10% slower than their non-atomic counterparts. This indicates that **atomic random memory accesses have similar rates to non-atomic ones on machines with newer CPU architectures**.

This finding suggests that performance optimizations aimed at preventing atomic accesses may have negative consequences, increasing execution time. Therefore, it is essential to consider the rates of different memory access types to fully leverage the capabilities of the processor architecture.

---

[2]https://prng.di.unimi.it/
[3]https://github.com/DIPSA-QUB/LaganLighter/blob/main/docs/5.0-random-mem-bench.md

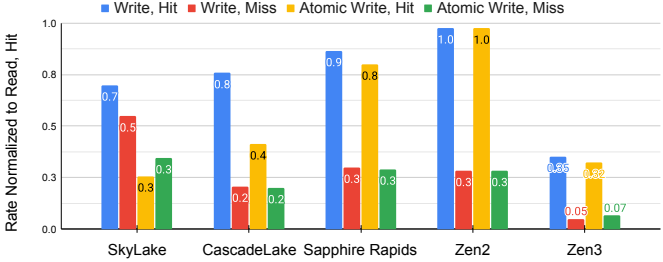

Fig. 3: Rate of random "Write" and "Atomic Write" accesses normalized to random "Read" access rate.

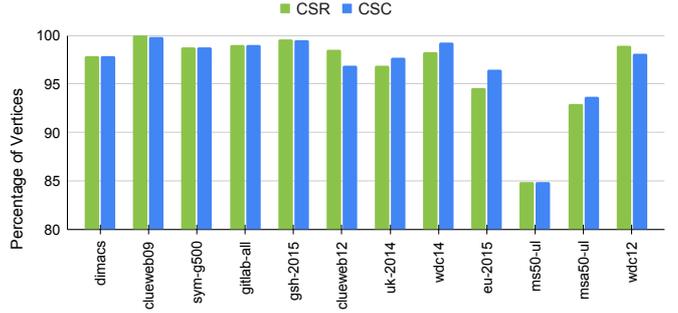

Fig. 4: Percentage of vertices with degrees less than 256

## C. Dataset Locality Impacts

The performance of graph analytics, such as GT, is dependent on locality of the input dataset. In a graph with high locality, a vertex and its neighbors are processed in a close time, increasing reuse of accessed data. For instance, in Atomic GT (Section II-B), the neighboring vertices are accessed in a short duration and their relevant indices in counters and IP arrays, are reused in cache. Therefore, one memory access is required for the first query and the subsequent queries are answered by the cache. In contrast, a low level of locality reduces cache reuse, leading to a poor performance. In other words, graph locality is the closeness of ID of neighboring vertices, which is reflected in the repetition of memory references when a block of consecutive vertices is traversed sequentially.

A graph dataset may have representations with different types of locality:

- **Natural Locality**: A graph dataset may inherently possess some levels of locality due to its creation or extraction procedure.
- **Optimized Locality**: Locality optimizing algorithms improve the locality of graphs [10]–[12].
- **Randomized Locality**: Relabeling of graph by randomized arrays degrades locality.

Several metrics have been proposed to measure graph locality: Graph Bandwidth, Average Gap Profile, Compressed Bits per Link [13], and Neighbor to Neighbor Average Id Difference [12].

Publicly available graph datasets have varying levels of locality. This makes it necessary to consider graph locality when selecting the best processing method. While complicated processes that improve cache reuse may benefit graphs with low locality, they may also have negative impacts on graphs with high locality levels by increasing processing overhead.

## D. Cache Space Usage Efficiency

Another approach to optimize cache locality is to reduce the size of each element intended to be cached [14], [15]. By doing so, a greater number of elements can be cached in the same cache space, resulting in reduced cache misses.

In the context of graphs with up to 4 billion vertices, each vertex ID requires dedicated 4 Bytes. Consequently, the counters array in Atomic GT (Section II-B) necessitates 32 bits per index. However, as Figure 4 shows, more than 84% of vertices have degrees less than 256. This implies that the most significant bytes of a large percentage of indices in the counters array are never accessed, yet they occupy valuable cache space. This presents an opportunity to **improve cache space utilization by dedicating cache to a greater number of smaller-sized elements**.

## IV. POTRA

### A. High-Level Algorithm

**Structure-Aware Memory Accesses**. The design of PoTra is based on the structure of real-world graphs, which determines how memory accesses are performed in GT. Real-world graphs typically follow a skewed degree distribution, characterized by a large number of Low-Degree Vertices (**LDV**) and a small number of High-Degree Vertices (**HDV**).

In a graph with a skewed degree distribution, HDV edges are frequent and incur memory accesses to indices of HDVs in the counters array, which comprise a small portion of the total indices. In contrast, the graph has a large number of LDV, and each LDV counter is referenced rarely. We leverage this feature to optimize cache utilization by **dedicating the cache space to HDV as they are few in number and frequently accessed**, resulting in the rapid processing of a large portion of edges (HDV edges). Section V-G shows that 1.5–73.5% (26%, on average) of edges are selected as HDV edges.

To accelerate the processing of HDV edges, PoTra assigns per-thread arrays to store the counters and IP data related to HDV. These arrays are **private to the thread** and accessed without atomic accesses. Contrary to HDV, LDV have shared arrays that are protected by atomic accesses. As a result, **PoTra requires a memory space proportional to the size of the hardware caches**. In contrast, previous works dedicate an array of O(|V|) size to each thread (Section III-A). This makes PoTra scalable to an increasing number of threads and imposes no limits to the size of the input graph.

**Identifying and Compressing HDV**. To identify the HDV of the transposed graph, PoTra performs a parallel **sampling** on the edges array of the input graph and identifies the top-repeated endpoints. Note that the HDV of the input graph can be identified by sorting the offsets array, but we need the HDV of the transposed graph which may not be the same.

To efficiently store HDV data in cache, it is required to keep the HDV in consecutive elements. However, the HDV are scattered throughout the range of vertex IDs. To address this, PoTra assigns a unique identifier, called *hdv_index*, to each HDV within a narrow range. We use a **hash table to look up the *hdv_index***. When processing a vertex in Step 1 or 3, PoTra checks the hash table to determine if the vertex is a HDV and to retrieve its `hdv_index`. If the vertex does not exist in the hash table, it is classified as a LDV.

**Number of HDV**. PoTra identifies the number of HDV based on (i) the cache size ($|cache|$), (ii) the required space in the hash table for each record ($|r|$), (iii) the desired load factor of the hash table ($\alpha$), (iv) the required space for data associated with each HDV (d) and (v) the number of threads (t). The number of HDV is denoted by **k** and is calculated as follows:

$$k = \frac{|cache|}{\frac{|r|}{\alpha} + d \cdot t} \quad (1)$$

**Data Size**. To optimize cache utilization and performance, we aim to minimize the amount of data associated to each LDV and HDV. As shown in Section III-D, a 1-Byte counter is enough for a large portion of the vertices as the collected counts do not exceed 255. Furthermore, the above formula indicates that by reducing the size of data required for each HDV, the number of HDV is increased. This, in turn, means that a greater portion of edges are processed as HDV, resulting in better coverage of HDV edges and improved performance.

To optimize space utilization in Step 1, we divide the `counters` array of HDV into two segments:

- **HDV_low_counters** contains the least significant bytes of the counters. The relevant index of this array is incremented when a HDV counter is incremented. Due to its size, we expect it remains in cache.
- **HDV_high_counters** contains the most significant bytes of the counter and is incremented after an overflow of the similar index in HDV_low_counters.

As HDV counters are private, each thread has its own *HDV_low_counters* and *HDV_high_counters*, each of size $k$ and is indexed by *hdv_index*.

In this subsection, we explained the main functionality of the PoTra which we refer to as the Hash-based LDV-HDV (**HLH**) method. Section IV-B compares the performance model of HLH vs. Atomic and Section IV-C explains the PoTra in details.

### B. Performance Model

Section III-B explained that the rate at which the hardware can execute random atomic memory accesses may be close to that for non-atomic accesses. In Section III-C we also saw that graphs may have different levels of locality, which may impact the effectiveness of locality optimizing methods. As such, in order to optimize GT, it is essential to investigate the relationship between the locality of the dataset and the rate of atomic and non-atomic memory accesses. To this end, we construct performance models for the Atomic GT and HLH

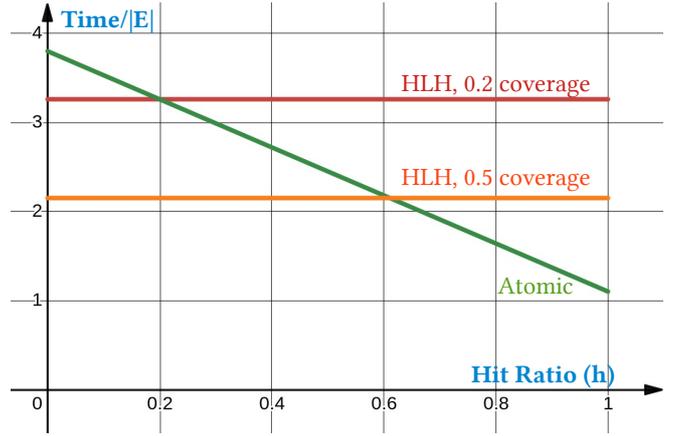

(a) Zen2

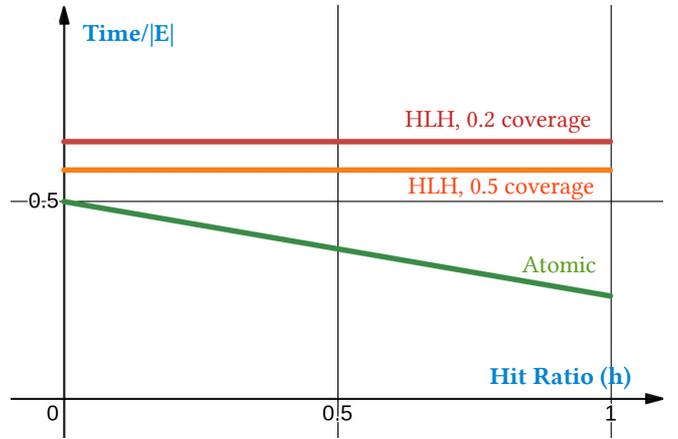

(b) Sapphire Rapids

Fig. 5: Performance model of Atomic vs HLH method

methods to determine when they are most effective.

**Atomic GT**: Let $h$ be the cache hit ratio for random accesses (which is affected by the locality of the input graph). The average time taken for random accesses per edge in Step 1 or Step 3 of Atomic GT (Algorithm 1), is:

$$\frac{T_{Atomic}}{|E|} = h\, t_{aw,h} + (1-h)t_{aw,m} = h(t_{aw,h} - t_{aw,m}) + t_{aw,m} \quad (2)$$

where $t_{aw,h}$ is the time for an atomic write access that is a cache hit and $t_{aw,m}$ is the time taken by an atomic write access that misses the cache.

**HLH**: Using HLH (explained in Section IV-A), the performance per edge is:

$$\frac{T_{HLH}}{|E|} = \frac{|HDV.E|t_{w,h} + |LDV.E|t_{aw,m} + |E|t_{r,h}}{|E|}$$
$$= \frac{|HDV.E|}{|E|}(t_{w,h} - t_{aw,m}) + t_{aw,m} + t_{r,h} \quad (3)$$

where $t_{w,h}$ is the time of a write memory access that is a cache hit. Also, $t_{r,h}$ is the time of a read memory access when it is a cache hit. $t_{r,h}$ indicates the time spent searching the hash table, assuming no collisions occur. $|HDV.E|$ and $|LDV.E|$ indicate the number of edges of HDV and LDV in the edges array of the input graph, respectively. $\frac{|HDV.E|}{|E|}$ is called **coverage**, indicating what fraction of the edges are pointing to HDV and are handled with on-chip counters by HLH.

Figure 5 compares the performance of Atomic and HLH for the Zen 2 and Sapphire Rapids machines (Section V-A shows details of machines). Each plot in Figure 5 includes the performance model for Atomic GT and two instantiation for the HLH model assuming 0.2 and 0.5 coverage is achieved. The coverage values of 0.2 and 0.5 are as we achieve, on average, 26% coverage on datasets (Section V-G).

Figure 5a shows that **for the Zen2 machine, and a graph with a high locality level, the Atomic method outperforms HLH**. However, **for a graph with a low locality level, HLH achieves better performance**. Additionally, Figure 5a demonstrates that increasing the coverage of edges in HLH provides more opportunities for performance improvement.

Figure 5b shows that **for the Sapphire Rapids machine, HLH with 0.2 and 0.5 coverage cannot outperform Atomic**. This is because $t_{aw,h}$ and $t_{aw,m}$ have close values, resulting in a flatter slope for $h$ in Equation 2. Consequently, HLH method requires a large coverage of about 0.8 to approach the performance of Atomic method with a small hit ratio. Section V-D includes a more detailed analysis.

The above discussion shows that the performance improvement of HLH largely depends on the locality of dataset and memory access timing. Therefore, **PoTra probes the performance of the HLH and Atomic methods for a small fraction of the graph edges** to determine which method performs better. The selected method is, then, used for the remainder of the edges.

### C. Detailed Algorithm

In this section, we provide a detailed explanation of the PoTra algorithm. Algorithm 2 presents the pseudo code.

**Step 0: Pre-processing**. As discussed in Section IV-B, the effectiveness of processing methods, depends on the locality of the graph. PoTra begins by evaluating the locality of the input graph and the cost of Atomic and HLH accesses.

The `potra_preprocessing()` (Line 1) performs a parallel sampling over the edges array to identify the HDV. The number of HDV, $k$, is determined by Equation 1. A hash table is then created to probe and compare performance of HLH against Atomic.

The sampling procedure involves randomly selecting 1% of $|V|$ number of samples from the edges array using parallel threads. To achieve this, the total range of the edges array and the number of samples are divided among threads. A temporary array is used to store the frequency of each vertex's appearance. This array is then sorted using the SAPCo sort algorithm [16] to identify the most frequent vertices, which are selected as HDV.

Next, PoTra determines which of Atomic and HLH are the most efficient algorithm for the current input graph and the

---

**Algorithm 2:** PoTra

**Input:** G(offsets, edges)

/* **Step 0: PoTra Pre-processing** */
1 (method, k, hash_map) = **potra_preprocessing**();
2 **if** method == "Atomic" **then**
3     **return** Atomic_GT(G);
4 counters[|G.V|] = {0};
5 HDV_low_cntrs[#threads][k] = {0};
6 HDV_high_cntrs[#threads][k] = {0};
7 part2tid[|G.partitions|];

/* **Step 1: Calculating Degrees** */
8 **par_for** p ∈ G.partitions
9     part2tid[p] = tid;
10     **for** e ∈ p.edges **do**
11        hdv_index = hash_map.find(e);
12        **if** hdv_index == -1 **then**
           /* A LDV edge */
13           fetch_and_add(counters[e], 1);
14        **else**
           /* A HDV edge */
15           val = ++HDV_low_cntrs[tid][hdv_index];
16           **if** val == 0 **then**
17             HDV_high_cntrs[tid][hdv_index]++;

/* **Step 2: Aggregating Counters** */
18 (t_offsets, IP, HDV_IP) = **potra_prefix_sum**();

/* **Step 3: Writing Edges** */
19 t_edges[|G.E|];
20 **par_for** tid ∈ threads
21     **for** p ∈ {G.partitions | part2tid[p] = tid} **do**
22        **for** v ∈ p.vertices **do**
23           **for** u ∈ v.neighbors **do**
24             hdv_index = hash_map.find(u);
25             **if** hdv_index == -1 **then**
                 /* A LDV edge */
26                 index = fetch_and_add(IP[u], 1);
27             **else**
                 /* A HDV edge */
28                 index = HDV_IP[tid][hdv_index]++;
29             t_edges[index] = v;
30 **return** (t_offsets, t_edges);

---

hardware it is executing on. Although we can experimentally measure the memory access rate for each type of memory access and use the performance model to predict, the hit ratio for the Atomic algorithm depends on the locality of the graph (and the order in which its vertices are labelled) and is not easily measured or predicted. Therefore, PoTra measures the execution time of HLH and Atomic for a small number of edges to identify the best method based on the input graph and memory access rates of the machine.

If the "Atomic" method is preferred (Line 2), PoTra calls Atomic GT to process the graph. Otherwise, PoTra continues processing the graph in HLH method. In Lines 4-7, the required arrays are initialized. In Step 1, the ID of partitions processed by a thread is stored in part2tid array, which will be used in Step 3 to ask each thread to process the partitions it has previously processed in Step 1. This is necessary as HDV counters and IP are private to threads, and the contents of these arrays need to line up between Step 1 and Step 3.

TABLE I: Machines

|  | SapphireRapids | Zen2 | Zen3 |
|---|---|---|---|
| Sockets | 2 | 2 | 2 |
| CPU | Intel 6438Y+ | AMD 7702 | AMD 7713 |
| Freq. | 0.8–4 GHz | 2.0–3.4 GHz | 2.0–3.6 GHz |
| Cores | 64 | 128 | 128 |
| Threads | 128 | 128 | 128 |
| L3 + L2 | 120 + 128 MB | 512 + 64 MB | 512 + 64 MB |
| Memory | 2 TB | 2 TB | 1 TB |
| OS | Debian 12 | CentOS 8 | Debian 12 |

TABLE II: Datasets

| Dataset | Type | \|V\| | \|E\| | Max. Out. | Max. In. |
|---|---|---|---|---|---|
| pkc | SN | 1.6M | 30.6M | 8.8k | 13.7k |
| en-wiki | WG | 6.6M | 165.2M | 12.4k | 236.5k |
| webbase-01 | WG | 118.1M | 1.0G | 3.8k | 816.1k |
| twitter-10 | SN | 41.7M | 1.5G | 3.0M | 770.2k |
| sk-2005 | WG | 50.6M | 1.9G | 12.9k | 8.6M |
| webcc | WG | 89.1M | 2.0G | 1.3M | 2.3M |
| ms1-ul | BG | 43.1M | 2.7G | 14.2k | 14.2k |
| friendster | SN | 65.6M | 3.6G | 5.2k | 5.2k |
| uk-union-07 | WG | 133.6M | 5.5G | 22.4k | 6.4M |
| dimacs | WG | 105.2M | 6.6G | 975.4k | 975.4k |
| clueweb09 | WG | 1.7G | 7.9G | 2.1k | 6.4M |
| g500 | SG | 536.9M | 17.0G | 3.8M | 3.8M |
| msa10-ul | BG | 1.8G | 25.2G | 207k | 62k |
| gitlab-all | VH | 1.1G | 27.9G | 387.7k | 19.4M |
| gsh-2015 | WG | 988.5M | 33.9G | 32.1k | 58.9M |
| clueweb12 | WG | 978.4M | 42.6G | 7.4k | 75.6M |
| uk-2014 | WG | 787.8M | 47.6G | 16.4k | 8.6M |
| wdc14 | WG | 1.7G | 64.4G | 32.8k | 45.7M |
| eu-2015 | WG | 1.1G | 91.8G | 35.3k | 20.3M |
| ms50-ul | BG | 585.6M | 124.8G | 507.8k | 507.8k |
| msa50-ul | BG | 1.8G | 125.3G | 543.1k | 298.0k |
| wdc12 | WG | 3.6G | 128.7G | 55.9k | 95.0M |

**Step 1: Counting degrees.** Lines 8-17 show how PoTra counts the degree of vertices. The graph is divided into partitions with an equal number of edges per partition. The partitions are dynamically allocated to threads. For each edge, the hash map is searched (Line 11) and based on the result, it is identified whether the endpoint is HDV or LDV. For LDV, the shared counters array is atomically incremented (Line 13) and for HDV, the private counter(s) are incremented (Lines 15-17).

**Step 2: Aggregating Counters.**

To identify the Insertion Points (IP) of vertices, all counters need to be aggregated. This is achieved by pass over vertices and querying the hash table. For a vertex $v$, if it is a LDV, *counters*[$v$] is added up. Otherwise, $v$ is a HDV and *HDV_high_cntrs*[*tid*][*hdv_index*] and *HDV_low_cntrs*[*tid*][*hdv_index*] of all threads are added up.

Parallelization of this process is similar to prefix_sum of Atomic GT. The vertices are divided into partitions and threads count the number of edges in each partition. A serial prefix sum is performed to identify the insertion point of the first vertex of each partition. A second parallel pass over vertices is required to specify the insertion point of all vertices (and all threads for HDV) in each partition.

**Step 3: Writing Edges.** Step 3 (Lines 20-29) is similar to Step 1. For LDVs, the shared IP is used to get the insertion point of the endpoint (Line 26). For HDVs, the hdv_index returned by the hash table is used to get the insertion point using the HDV_IP[tid] (Line 28).

## V. EVALUATION

### A. Experimental Setup

We present experiments on three processor architectures listed in Table I. The graph datasets are shown in Table II, along with their maximum in-degree and out-degree. Graph types include: Web Graphs (WG) [3], [5], [13], [17]–[19], Social Networks (SN) [20], [21], Synthetic Graph (SG) [22], [23], Version Control History Graphs (VH) [1], [4], and Bio Graphs (BG) [2], [24], [25].

As explained in Section III-C, a dataset may have three locality types: natural, optimized, and randomized. The web graph datasets are published with natural and optimized locality. The bio graphs are published in natural order. Additionally, the datasets can be in CSR or CSC direction (Section II-A). Converting from CSR to CSC may have different performance than converting from CSC to CSR due to discrepancies in the in-degree and out-degree distributions, as well as differences in locality in either direction [12].

To better understand the relationship between locality, direction, and performance, we evaluate each dataset in 4 representations:

- **CSR**: CSR with optimized (preferred) or natural locality,
- **CSR Rnd.**: CSR with randomized locality,
- **CSC**: CSC with optimized (preferred) or natural locality,
- **CSC Rnd.**: CSC with randomized locality.

We use Atomic GT as the baseline for comparison between PoTra and previous works. Table III shows the performance of Atomic GT on Zen2 machine for different representations of graphs. It clearly indicates that graph transposition is a non-trivial and time-consuming operation. Notably, the CSR and CSC representations of web graphs have optimized locality, resulting in shorter execution times compared to graphs of similar sizes.

In this section, we do not include the sort time in experiments, except in Section V-B, where the execution of all algorithms produce sorted results.

We implemented PoTra in the C language using Lagan-Lighter [26], ParaGrapher [27], OpenMP [28], and libnuma. Code is compiled with gcc-9.2 using the -O3 flag.

TABLE III: [Zen2] Performance of Atomic GT in seconds

| Dataset | |V| | |E| | CSR | CSR Rnd. | CSC | CSC Rnd. |
|---|---|---|---|---|---|---|
| webbase-01 | 118.1M | 1.0G | 0.338 | 3.2 | 0.281 | 3.9 |
| twitter-10 | 41.7M | 1.5G | 2.7 | 3.7 | 5.8 | 3.3 |
| sk-2005 | 50.6M | 1.9G | 1.7 | 3.3 | 0.816 | 6.5 |
| webcc | 89.1M | 2.0G | 2.6 | 4.9 | 2.6 | 5.3 |
| ms1-ul | 43.1M | 2.7G | 5.7 | 7.7 | 5.8 | 7.5 |
| friendster | 65.6M | 3.6G | 11.1 | 13.8 | 10.9 | 13.7 |
| uk-union-07 | 133.6M | 5.5G | 3.3 | 13.2 | 1.6 | 29.7 |
| dimacs | 105.2M | 6.6G | 1.5 | 24.3 | 1.5 | 23.8 |
| clueweb09 | 1.7G | 7.9G | 14.7 | 85.1 | 21.5 | 96.8 |
| g500 | 536.9M | 17.0G | 116.4 | 115.7 | 116.8 | 116.3 |
| msa10-ul | 1.8G | 25.2G | 179.8 | 275.3 | 191.6 | 275.6 |
| gitlab-all | 1.1G | 27.9G | 9.5 | 212.6 | 11.4 | 245.8 |
| gsh-2015 | 988.5M | 33.9G | 11.5 | 302.5 | 8.3 | 327.7 |
| clueweb12 | 978.4M | 42.6G | 103.4 | 345 | 16.2 | 390.6 |
| uk-2014 | 787.8M | 47.6G | 11.3 | 392.3 | 9.3 | 449.1 |
| wdc14 | 1.7G | 64.4G | 17.6 | 315.7 | 34.7 | 763.4 |
| eu-2015 | 1.1G | 91.8G | 16 | 802.7 | 18.2 | 897.8 |
| ms50-ul | 585.6M | 124.8G | 1039.5 | 1122.3 | 1038 | 1119.2 |
| msa50-ul | 1.8G | 125.3G | 1069.1 | 1410.2 | 1159.5 | 1419.6 |
| wdc12 | 3.6G | 128.7G | 47.6 | 1350 | 42.7 | 1496.4 |

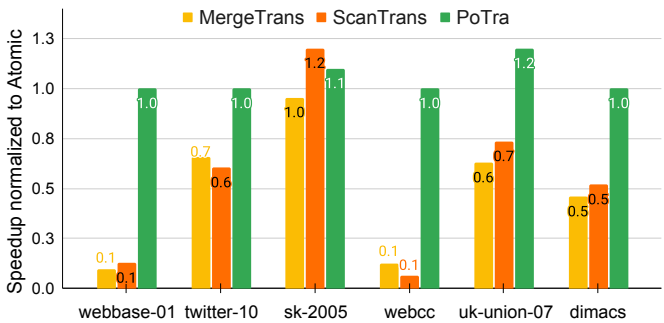

Fig. 6: [Zen2, CSR] Speedup of ScanTrans, MergeTrans, and PoTra normalized to Atomic GT.

### B. Evaluation of Previous Works

Figure 6 compares the performance of ScanTrans, MergeTrans, and PoTra on the Zen2 machine. The values are normalized to the execution time of Atomic GT (Table III). As explained in Section III-A, ScanTrans and MergeTrans require memory memory space of O(#threads·|V|). This leads to failed execution with an "Out of Memory" error for graphs with large orders such as clueweb09. Therefore, we report results for graphs with sizes of up to dimacs.

Figure 6 reveals that, **except for the sk-2005 graph, MergeTrans and ScanTrans have a lower performance than both PoTra and Atomic GT**. Additionally, Figure 6 reveals that MergeTrans and ScanTrans perform at worst for graphs with larger |V|, such as webbase-01 and webcc, due to the increased cost of aggregating counters of threads in Step 2, which is dependent on |V|.

### C. Zen Machines

Figure 7 compares the speedup of PoTra over Atomic on the Zen2 and Zen3 machines for CSR and CSR Rnd. representations of the small graphs in our evaluation. Since the number of vertices is small (|V| < 200M), a large fraction of the counters array in Step 1 and also IP array in Step 3 is easily kept in cache. As a result, Atomic GT is executed with high locality, and the HLH improvements are limited to up to 30%. A similar trend is experienced in CSC and CSC Rnd. formats of small graphs, but these results are not shown.

Figure 8 shows the speedup of PoTra for CSR and CSR Rnd. representations of the graphs larger than dimacs on the Zen2 and Zen3 machines. The CSC and CSC Rnd. representations are shown in Figure 9. The plots indicate that PoTra provides a speedup of 0.9–6.2 times on Zen2 and 0.9–8.7 times on Zen3. **The average speedup on Zen2 and Zen3 is 1.6, 1.7 times, respectively**.

PoTra performs a probing step to select Atomic or HLH (Step 0, Algorithm 2). If Atomic is preferred, the overhead of probing is imposed, resulting in a 10% performance loss compared to Atomic GT. However, this mainly occurs for graphs with optimized locality (with small execution times) and the overhead remains limited to a few seconds.

For some graphs with optimized locality (such as gitlab-all, gsh-2015, and wdc14), we do not see performance improvement for CSR and CSC representations (i.e., 0.9–1.0 times speedup). This indicates that PoTra has selected to invoke Atomic algorithm, as chosen by the probing procedure. In contrast, for clueweb12, we see that HLH has been able to facilitate 6.2 times speedup on Zen2 and 8.7 times on Zen3. The clueweb12 graph comes with optimized locality, showing that **PoTra can improve performance of GT also for datasets with optimized locality**.

On the other hand, for randomized graphs with low locality levels, PoTra tries to keep the data of HDV as frequently-referenced vertices in cache. By allocating private data to each thread, PoTra prevents cache contention and costs of cacheline migration. In this way, we observe that **PoTra has provided 1.4–3.5 times speedup for randomized graphs**.

### D. Sapphire Rapids Machine

In Section IV-B, we discussed the impacts of machine architecture and graph locality on the effectiveness of PoTra. Figure 5a shows that PoTra works on the Zen2 machine as the cost of hash table search is offset by the time saved for avoiding memory accesses missed by cache. In contrast, Figure 5b suggests that HLH needs a large coverage to be effective on Sapphire Rapids. Our experimental evaluation confirms this, with **HLH achieving a speedup of 0.5–3.3 times, averaging 0.9±0.3 times on the Sapphire Rapids machine**. We investigate the subject in more details.

As shown in Table I, we utilized hyperthreading on the Sapphire Rapids machine as it delivers an average acceleration of 40% for both HLH and Atomic. The machine operates at a frequency of 2.8 GHz when deploying 128 threads, resulting in a CPU cycle time of approximately 350 picoseconds (ps).

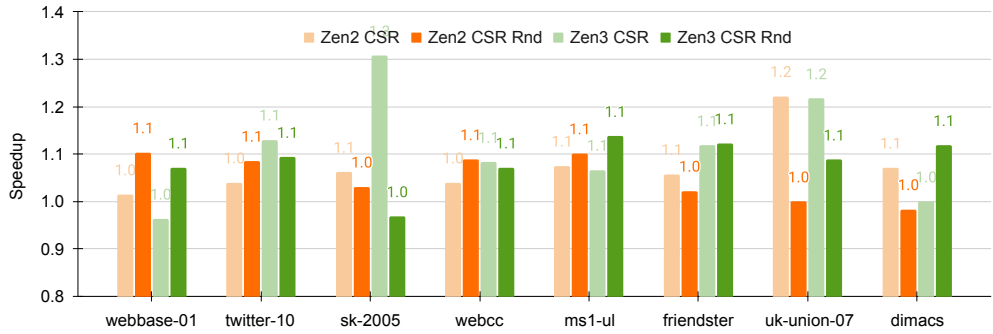

Fig. 7: [Zen2-3, CSR and CSR Rnd.] Speedup of PoTra over Atomic GT.

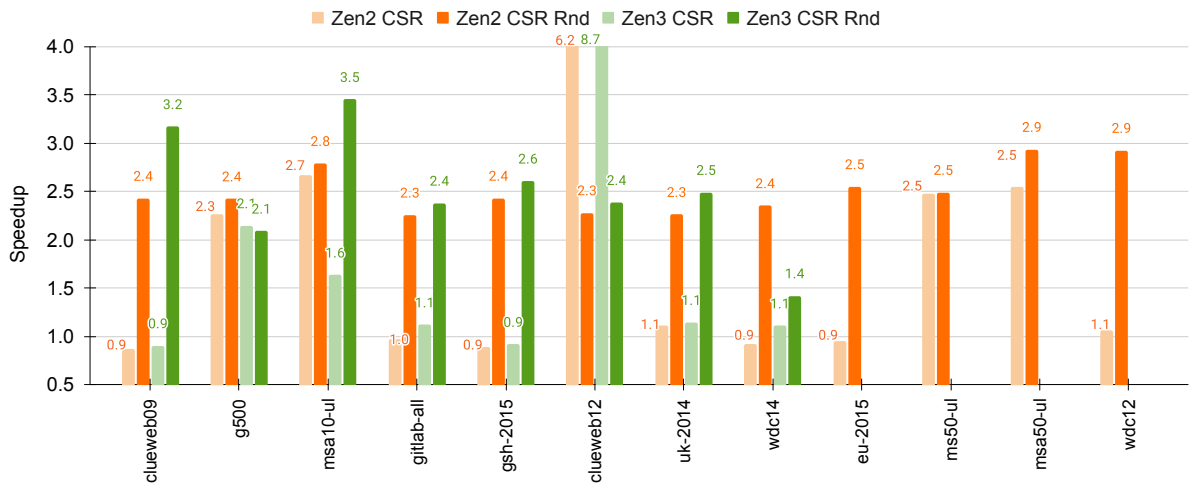

Fig. 8: [Zen2-3, CSR and CSR Rnd.] Speedup of PoTra over Atomic GT.

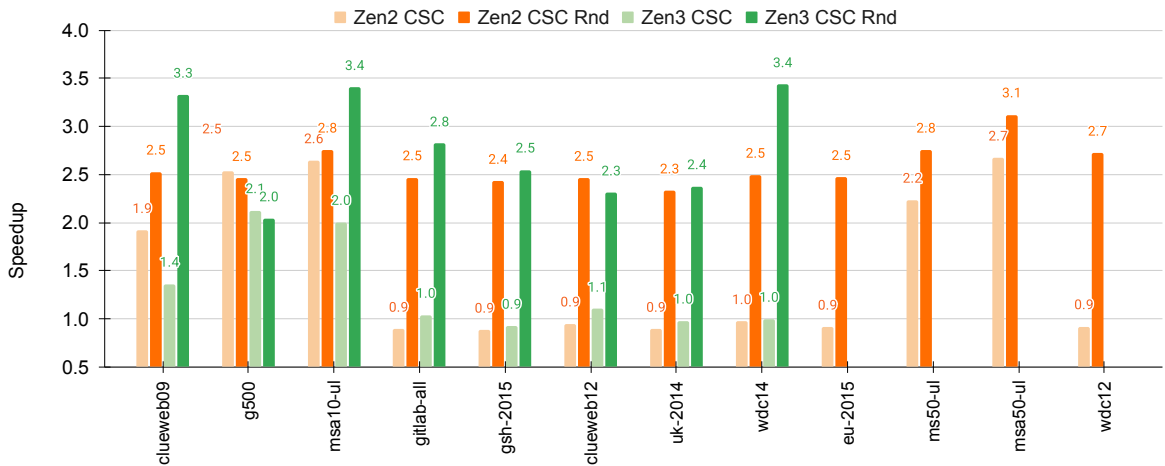

Fig. 9: [Zen2-3, CSC and CSC Rnd.] Speedup of PoTra over Atomic GT.

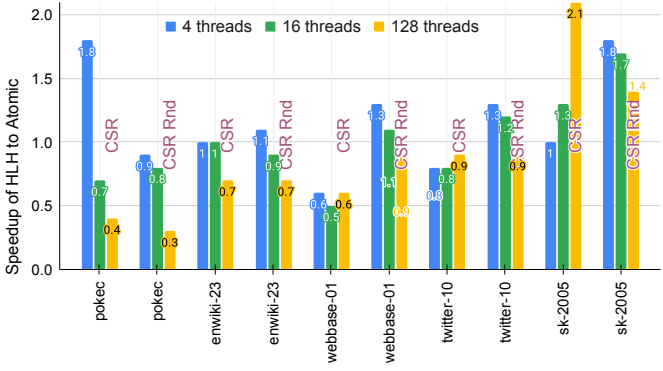

Fig. 10: [Sapphire Rapids, CSR and CSR Rnd.] Speedup of HLH compared to Atomic

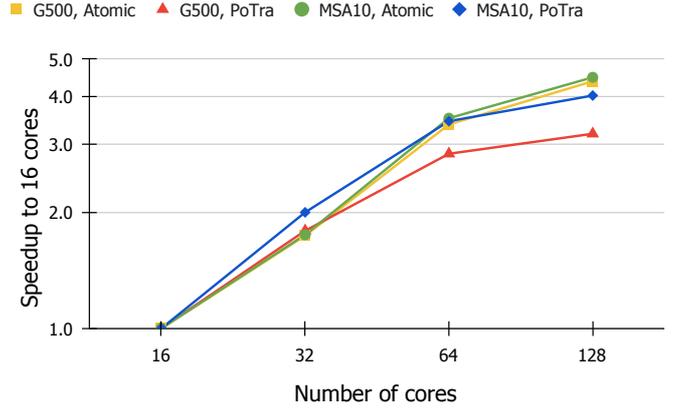

Fig. 11: [Zen2, CSR] Speedup of PoTra and Atomic

However, as Equation 3 indicates, the effectiveness of HLH depends on the difference of between $t_{aw,m}$ and $t_{w,h}$, which are 540 ps and 260 ps, respectively. This difference is shorter than a CPU cycle. In other words, the sum of a hit write access (for updating the cached private counter) and a CPU cycle (a minimal cost of searching the hash table) exceeds the time required for a missed atomic write access (in Atomic GT). Consequently, **at a frequency of 2.8 GHZ, the cost of hash table lookup cannot be compensated by the gain of avoiding missed access**.

To further investigate this, we varied the CPU frequency by changing the number of threads to 4 and 16 threads and measured the speedup of HLH compared to Atomic. The results are presented in Figure 10. Notably:

- The processor distributes its power budget across active cores, allowing a small number of cores to operate at a higher frequency. In our evaluation, the frequency is 4 GHz for 4 and 16 threads.
- Reducing the number of threads increases $t_{aw,m}$ and $t_{w,h}$. For 16 threads, $t_{aw,m}$ = 1.6 ns, $t_{w,h}$ = 0.6 ns and for 4 threads, $t_{aw,m}$ = 5.9 ns, $t_{w,h}$ = 2.6 ns.
- By reducing the number of threads, PoTra benefits from a greater coverage as $k$ is increased ( Equation 1).

In this way, by reducing the number of threads, the frequency is increased and the processor has sufficient clock cycles between $t_{aw,m}$ and $t_{w,h}$ to allocate to searching the hash table. As a result, the cost of HLH is offset by the speedup it provides. Figure 10 also demonstrates that reducing the number of threads typically leads to an increase in the speedup of HLH, particularly, for the randomized representation up to 1.8 times speedup.

*E. Scalability*

Figure 11 compares execution of PoTra and Atomic for different number of cores, with results normalized to the execution time of 16 cores. In this measurement, both PoTra and Atomic were restricted to using only the memory nodes and CPU caches available to the active cores. The Zen2 machine consists of 8 NUMA nodes, each comprising of 16 cores and 4 L3 caches, each with a size of 16 MB.

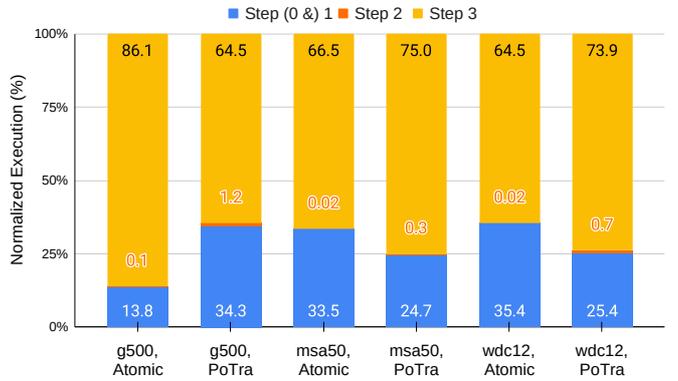

Fig. 12: [Zen2, CSR Rnd.] Execution time decomposition

The results in Figure 11 demonstrate that both Potra and Atomic exhibit competitive single-machine scalability. However, when increasing the number of cores from 16 to 128 (an 8-fold increase), we observe only 3.2- and 4-fold speedups. Across all cores and graphs measured in this experiment, PoTra delivers a 72.8% parallel efficiency, on average.

This sublinear scaling may be attributed to various factors, including reduced frequency, increased contention in the L3 caches, a non-linear increase in memory bandwidth, and potential issues in memory allocation procedure, such as unbalanced or non-local memory allocation on NUMA nodes for graph topology and vertex data.

*F. Execution Details*

**Execution Decomposition**. Figure 12 compares the percentage of time spent in the execution of Atomic and PoTra for the CSR Rnd representation of several graphs. We have selected the graphs for which PoTra chooses HLH in probing procedure. Figure 12 shows that PoTra and Atomic spend most of their time in Step 3 (placing the transposed edges). However, PoTra spends a larger proportion of its execution time in Step 2 as it needs to aggregate the private counters of threads and produce their IP. Notably, Step 2 remains a relatively insignificant part of the algorithms.

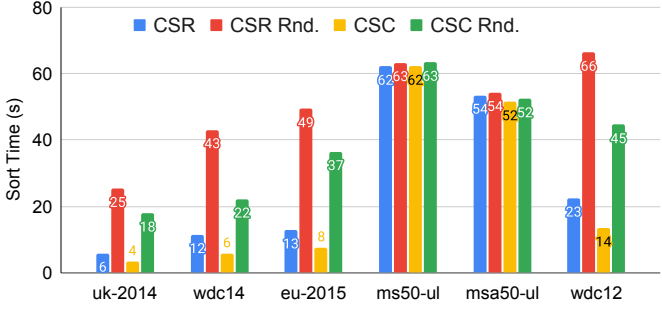

Fig. 13: [Zen2] Execution time of sorting

TABLE IV: [Zen2] Sampling time and coverage

| Dataset | Direction | Time(s) | Coverage% |
|---|---|---|---|
| g500 | CSR | 0.4 | 39.6 |
| g500 | CSC | 0.4 | 39.6 |
| clueweb12 | CSR | 0.4 | 22.2 |
| clueweb12 | CSC | 0.4 | 0.5 |
| wdc14 | CSR | 0.7 | 70.6 |
| wdc14 | CSC | 0.7 | 0.6 |
| eu-2015 | CSR | 0.7 | 21.4 |
| eu-2015 | CSC | 0.7 | 1.6 |
| msa50-ul | CSR | 0.8 | 2.7 |
| msa50-ul | CSC | 0.8 | 1.7 |
| wdc12 | CSR | 1.5 | 23.5 |
| wdc12 | CSC | 1.6 | 1.7 |

**Step 3 Load Imbalance**. As explained in Section IV-C, each thread in Step 3 processes the same partition of edges it had processed in Step 1, resulting in a load imbalance between threads in Step 3. Our evaluation on Zen2 machine shows that Step 3 has a load imbalance of 3.8±4.4%. A potential solution to this issue could be to perform Step 3 in two phases:

- Phase 1: Each thread processes 80% of its partitions (i.e., the partitions it processed in Step 1),
- Phase 2: Threads loop over partitions and *atomically*[4] process the remaining partitions of those threads that have completed Phase 1.

We have not implemented this solution.

**Sorting**. The sort time of Atomic and PoTra are almost identical. To illustrate the cost of sorting, Figure 13 compares the sort time of four representations for the largest graphs. A serial quick sort has been used for sorting neighbor lists of each vertex and the work was parallelized over vertices. Figure 13 shows that **the randomized representations have longer sorting times**.

### G. Sampling

Table IV presents the execution time and coverage of the top 1 million selected HDV for graphs in CSR and

[4]Atomic increment of *HDV_IP* (Algorithm 2, Line 28).

CSC representations. The results are the same on randomized representations. PoTra samples 1% of |V|, however, an upper bound for the number of samples can be determined based on the frequency of elements [29]. Note that web graphs usually do not have large max. out-degrees (Table II), which results in small coverage rates for their CSC representations.

### H. Energy Reduction

Figure 14 illustrates the speedup and energy reduction of PoTra in comparison to Atomic for the Zen2 and Zen3 machines and for the graphs in CSR and CSR Rnd. representations. In measuring energy consumption, we utilized the processor's energy counters, which account for enery consumed by processors but exclude energy consumed by memory modules. We observe a relative difference of 8.3 ± 8.1% between speedup and energy reduction, indicating that the energy reduction values are close to speedup values.

## VI. FURTHER RELATED WORKS

**Transposition**. Gustavson proposes a transposition method that treats edges as ordered sets, sorting them by their second endpoint, and then compressing the format [30]. This approach requires additional memory space of $(2 \cdot |E|)$ and involves multiple passes ($O(log(|E|))$) over the uncompressed edges to perform the sort. Gonzalez-Mesa et al. investigate the use of transactional memory for a GT similar to MergeTrans [31]. In-memory and out-of-core transposition of dense matrices have been studied in [32]–[34].

**Locality Optimizing**. Various locality-optimizing algorithms have been suggested [10]–[12], [35], [36]. These algorithms use heuristics such as community detection or simulating cache to assign close IDs to neighboring vertices, thereby improving locality.

**Structure-Aware Algorithms**. Researchers have leveraged graph structure to enhance load balance [37], reduce memory accesses [38], [39], improve locality of memory accesses [26], [40], [41], increase work efficiency [16], and to provide better partitioning [42], [43]. SDS Sort is a parallel sorting algorithm designed for data with skewed distribution [44]. SAPCo Sort [16] is an optimized degree-ordering for graphs with skewed degree distributions.

**Architecture-Aware Optimizations**. The effectiveness of cache-oblivious and cache-conscious methods for sorting [45] and linear algebra [46] has been studied. NUMA-aware algorithms for graph analytics have been developed [47], [48]. Techniques such as cache partitioning and frequency scaling have been used to reduce energy consumption in heterogeneous multi-core architectures [49]. Architecture specifications have also been used to optimize precision [50], prune deep neural networks [51], and improve virtual memory performance [52].

## VII. CONCLUSION

In this paper, we investigate the impact of graph structure and graph locality on memory accesses in graph transposition, a widely-used and time-consuming graph algorithm.

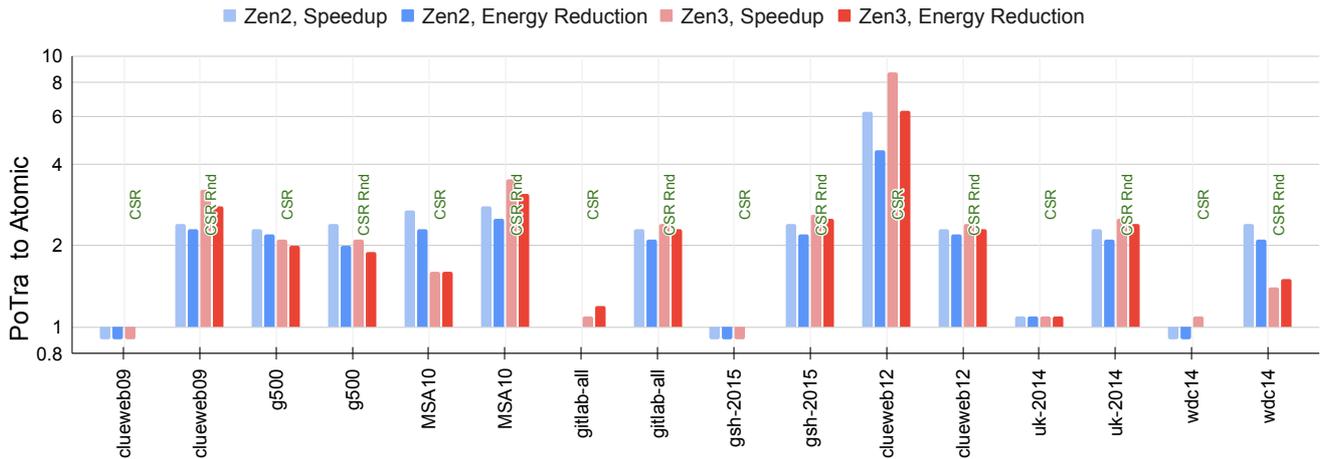

Fig. 14: [Zen2,Zen3, CSR, CSR Rnd.] Speedup and energy reduction of PoTra compared to Atomic

We designed PoTra, a structure- and architecture-aware graph transposition. By modeling its performance, we studied the dependency of PoTra on memory access rates. Our evaluation on 20 real-world graphs with up to 128 billion edges demonstrates that PoTra achieves up to 8.7 times speedup.

## SOURCE CODE

The source code of PoTra is available online on https://blogs.qub.ac.uk/DIPSA/LaganLighter/.

## ACKNOWLEDGMENT

This work is partially supported by the Engineering and Physical Sciences Research Council [EP/Z531054/1] (the Kelvin Living Lab), [EP/T022175/1] (Kelvin-2 Tier-2 HPC), [EP/X01794X/1] (ASCCED) and [EP/X029174/1] (RELAX). and by the Department for the Economy, Northern Ireland under grant agreement USI-226.